**Tailoring the electronic transitions of NdNiO$_3$ films through (111)$_{pc}$ oriented interfaces**

S. Catalano,[1,a)] M. Gibert,[1] V. Bisogni,[2,b)] F. He,[3] R. Sutarto,[3] M. Viret,[1,4] P. Zubko,[1,c)] R. Scherwitzl,[1] G. A. Sawatzky,[5] T. Schmitt,[2] J.-M. Triscone[1]

[1]*Department of Quantum Matter Physics, University of Geneva, CH-1211 Geneva, Switzerland*
[2]*Swiss Light Source, Paul Scherrer Institut, CH-5232, Villigen, Switzerland*
[3]*Canadian Light Source, Saskatoon, Saskatchewan, S7N 2V3, Canada*
[4]*Service de Physique de l'Etat Condensé (CNRS UMR 3680), CEA Saclay, 91191, Gif-sur-Yvette, France*
[5]*Department of Physics and Astronomy, University of British Columbia, Vancouver, British Columbia, V6T 1Z1, Canada*

Bulk NdNiO$_3$ and thin films grown along the *pseudocubic* (001)$_{pc}$ axis display a 1$^{st}$ order metal to insulator transition (MIT) together with a Néel transition at T=200K. Here, we show that for NdNiO$_3$ films deposited on (111)$_{pc}$ NdGaO$_3$ the MIT occurs at T=335K and the Néel transition at T=230 K. By comparing transport and magnetic properties of layers grown on substrates with different symmetries and lattice parameters, we demonstrate a particularly large tuning when the epitaxy is realized on (111)$_{pc}$ surfaces. We attribute this effect to the specific lattice matching conditions imposed along this direction when using orthorhombic substrates.

a) Author to whom correspondence should be addressed. Electronic mail: sara.catalano@unige.ch
b) Currently at National Synchrotron Light Source II, Brookhaven National Laboratory, Upton, New York, 11973, USA.
c) Currently at London Centre for Nanotechnology and Department of Physics and Astronomy, University College London, 17-19 Gordon Street, London WC1H 0HA, UK.

Achieving control of the electronic phases of ABO$_3$ perovskites is today a major challenge for succeeding in the design of a new generation of functional devices.[1,2] In high quality epitaxial oxide films, exceptional properties can be established through interface effects, dimensionality control and heterostructuring.[3-5] In particular, by selecting the film/substrate lattice misfit, one can tune the *B-O* bond lengths and the *B-O-B* bond angles (Θ), modifying the crystal field and bandwidth of perovskite compounds.[6,7] The effect of epitaxial strain ($\varepsilon_{xx}$), induced by the film/substrate lattice parameter mismatch, has been extensively investigated in the past.[2,5,8] Both theoretical and experimental studies have suggested that, beyond pure strain effects, substrate symmetry and orientation, which can modify the distortions and tilting of the oxygen octahedra network, play an important role and are key to understanding the complex properties of oxide heterostructures.[9-14]

NdNiO$_3$ (NNO) is a member of the perovskite nickelate family (*R*-NiO$_3$, *R*=Rare Earth), whose phase diagram is controlled by the *Ni-O-Ni* bond angles.[15-18] In its bulk form, NNO undergoes a first order metal-to-insulator transition (MIT) that occurs along with a paramagnetic (PM) to antiferromagnetic (AFM) Néel transition at T$_{MI}$ = T$_{Néel}$ = 200 K. The AFM ground state is identified by the Bragg vector q$_{Bragg}$ = (¼ ¼ ¼) in *pseudocubic* (*pc*) notation.[16,17,19] By applying different levels of epitaxial strain on (001)$_{pc}$ oriented films, the occurrence of the MI and Néel transitions can be tuned over a wide temperature range,[20-25] the two transitions always occurring simultaneously, i.e. $T_{MI} = T_{Néel}$. A recent work revealed that NNO films, grown along other crystallographic axes on top of orthorhombic NdGaO$_3$, display a MIT at exceptionally high temperature.[26] The authors suggest that the strong in-plane anisotropy, characterizing the orthorhombic symmetry, plays a role in controlling the film properties.[26] Finally, it is worth mentioning that a conducting AFM ground state can be stabilized in PrNiO$_3$ films by confinement and epitaxial strain.[27]

Here, we investigate heterointerface effects by depositing NNO films on top of (001)$_{pc}$ and (111)$_{pc}$ oriented substrates with different symmetries and lattice parameters. By performing transport and resonant soft X-Ray diffraction measurements, we directly probe both the conduction and magnetic properties of the films. Our observations show that, by selecting substrates presenting an orthorhombic distortion and the (111)$_{pc}$ crystallographic direction for the growth, one can induce a splitting of $T_{MI}$ and $T_{Néel}$ over a temperature range never achieved before for this compound. We attribute this result to a sizeable change of the *Ni-O-Ni* bond angles in the film, induced by the specific lattice matching conditions imposed at (111)$_{pc}$ oriented perovskite interfaces.

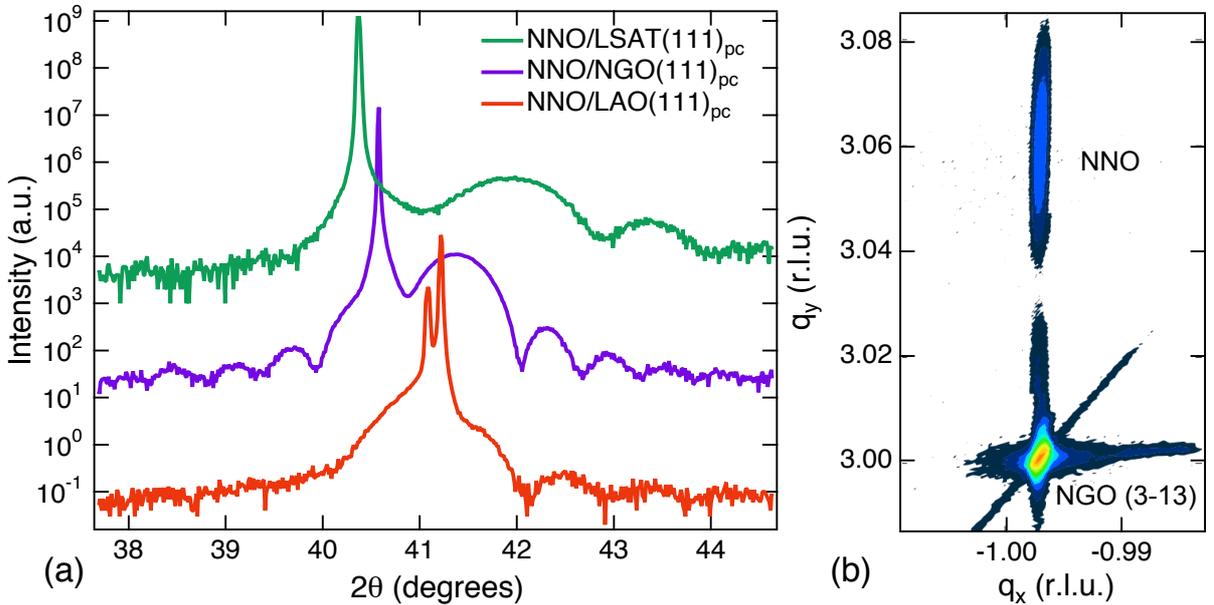

**Figure 1:** Characterization of the (111)$_{pc}$-NNO films using X-Ray diffraction. (a) Typical ω−2θ scans of 10 nm NNO/LSAT (green), 10 nm NNO/LAO (red), and 17 nm NNO/NGO (blue) films. (b) Reciprocal Space Map showing the 17 nm NNO film lattice parameter distribution around the (3 -1 3) orthorhombic reflection of (111)$_{pc}$ NGO, with the q$_y$ axis oriented along the (101)$_{ortho}$ out-of-plane direction and q$_x$ along (010)$_{ortho}$ in-plane. The film appears to be fully strained to the substrate.

Fully strained NNO films were deposited on top of (001)$_{pc}$ and (111)$_{pc}$ oriented single crystals of (LaAlO$_3$)$_{0.3}$(Sr$_2$AlTaO$_6$)$_{0.7}$ (LSAT), NdGaO$_3$ (NGO) and LaAlO$_3$ (LAO) using deposition conditions described previously.[22,23] The chosen substrates impose distinct levels of epitaxial strain and symmetry misfit. LSAT is cubic (space group *Pm3m*), NGO is orthorhombic (space group *Pbnm*) and LAO is rhombohedral (space group *R3c*). The applied strain values are $\varepsilon_{xx}$ = +1.6% for NNO/LSAT, $\varepsilon_{xx}$ = +1.5% for NNO/NGO, and $\varepsilon_{xx}$ = -0.5% for NNO/LAO. Here, $\varepsilon_{xx}$ is defined as $\varepsilon_{xx} = (a_{sub} - a_{NNO})/a_{NNO}$, $a_{sub}$ and $a_{NNO}$ corresponding to the *pc* lattice constant of the substrate and of bulk NNO. Note that the (001) *pseudocubic* crystallographic plane corresponds to the (110) plane in orthorhombic notation and the (111)$_{pc}$ to (101)$_{ortho}$. To allow for an easy comparison of equivalent crystallographic directions, *pseudocubic* notations will be used in the following.

Sample characterization was performed *ex-situ* with standard X-Ray diffraction techniques (using Cu K-$\alpha_1$ radiation, $\lambda$ = 1.5406 Å). Figure 1 displays examples of $\omega$–$2\theta$ scans (a) and a Reciprocal Space Map (b) for $(111)_{pc}$-oriented films which confirm the excellent crystallinity of the samples. The quality of $(001)_{pc}$ films was reported in a previous work.[22] Film thicknesses were kept between 10 nm and 17 nm, a thickness range for which films are fully strained to the substrate lattice as confirmed by reciprocal space maps (see an example in Figure 1b). The surface topography, studied by atomic force microscopy, reveals that the average roughness is lower than 5 Å.[28]

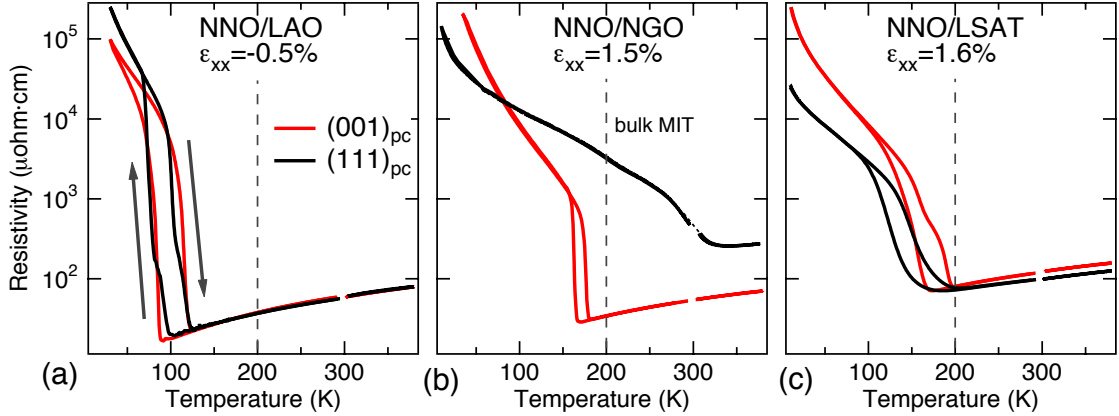

**Figure 2**: Resistivity *versus* temperature of NNO films grown along both $(001)_{pc}$ (red) and $(111)_{pc}$ (black) crystallographic directions on (a) LAO, (b) NGO and (c) LSAT. All films show a 1$^{st}$ order MIT with the exception of $(111)_{pc}$ NNO/NGO which displays a MI transition at 335 K with no hysteresis. The vertical dashed line indicates the bulk $T_{MI}$. The discontinuity of the data close to 300 K is due to a change in the measurement set up. Note the arrows in (a), specifying the cooling (upward arrow) and heating (downward arrow) temperature cycle.

Figure 2 shows the resistivity *versus* temperature of the NNO films of Figure 1 compared to the corresponding $(001)_{pc}$ oriented ones in the temperature range 500 K-4 K. The transport measurements were performed using a standard 4 probes configuration. A custom-made dipping station was used for measurements below 300 K. For measurements above room temperature, heating and cooling cycles were controlled using Peltier elements. $T_{MI}$ is defined as the temperature at which, for the cooling cycle, $-\partial \ln(R)/\partial T$ has a maximum for MI transitions showing hysteresis and as the temperature at which the resistance *versus* temperature curve displays a kink for the other cases.[23] All films undergo a MIT as the temperature is lowered. Both $(001)_{pc}$- and $(111)_{pc}$-oriented NNO films grown on top of LSAT and LAO (Figure 2a and 2c) display a sharp MIT at, respectively, $T_{MI}$ = 170 K and $T_{MI}$ = 100 K. Consistently with previously reported results, the MIT occurs with a visible hysteresis at a critical temperature $T_{MI}$ lower than in bulk ($T_{MI}$-bulk is indicated by the dashed line in Figure 2) and $T_{MI}$ decreases as the compressive strain increases.[20,22,24,29]

The transport characterization of NNO/NGO (Figure 2b), however, reveals striking differences depending on the substrate orientation. $(001)_{pc}$ NNO/NGO undergoes a 1$^{st}$ order MIT at about 160 K, a behavior rather similar to that of NNO/LAO and NNO/LSAT films. Surprisingly, the 17 nm thick film grown on $(111)_{pc}$ NGO exhibits a MIT at $T_{MI}$ = 335 K, with no detectable sign of hysteresis. In the metallic phase, the resistivity of this latter film is higher (by a factor of 2) than the values measured for the other cases. Also, as can be seen on Figure 2b, in the insulating phase, the resistance rises with a lower $\Delta R/\Delta T$ rate as compared to the other films. Clearly, these results point to a non trivial effect of the epitaxial mechanism operating at the $(111)_{pc}$ NNO/NGO interface – that is pushing the film-MIT 135 K above the MI and Néel transitions of bulk NNO.

In order to further probe the behavior of the $(111)_{pc}$ NNO/NGO films, we studied their magnetic properties. The onset of the AFM phase of $R$NiO$_3$ was investigated by resonant soft X-Ray reflectivity measurements carried out at the Canadian Light Source of Saskatoon (CLS)[30]. The Bragg vector $q_{Bragg} = (¼ ¼ ¼)$ could be accessed in specular geometry for $(111)_{pc}$ oriented films, whereas a special diffraction set up was used for the $(001)_{pc}$ oriented films.[23] The incoming light was tuned to the Ni $L_3$-edge (853 eV) and both linearly polarized in-plane (σ) and out-of-plane (π) light were used. Figure 3a shows the symmetric ω–2θ scans performed with π polarized light for temperatures from 380 K to 20 K for a 17 nm $(111)_{pc}$ NNO/NGO film. The diffractogram displays sharp finite size oscillations, corresponding to the film thickness. As the temperature is reduced, a strong superstructure reflection emerges at a value of 2θ corresponding to $q_{Bragg} = (¼ ¼ ¼)_{pc}$, matching the periodicity of the particular AFM ordering of the nickelates.[16,21,23,31-34] The signal also exhibits strong linear dichroism (inset of Figure 3a), which points to the magnetic origin of the resonant scattering mechanism[33]. This peak can therefore be safely ascribed to the antiferromagnetically-ordered phase of $R$-NiO$_3$. Figure 3b shows the temperature dependence of the intensity of the Bragg reflection (red curve, left axis) and of the resistivity (grey curve, right axis). These data show that an AFM-ordered insulating ground state develops in the $(111)_{pc}$ NNO/NGO films at $T_{Néel}$ = 230 K, while $T_{MI}$ is about 335 K. To our knowledge, these properties are unprecedented for NNO and spotlight a unique effect of the lattice matching for $(111)_{pc}$ NNO/NGO.

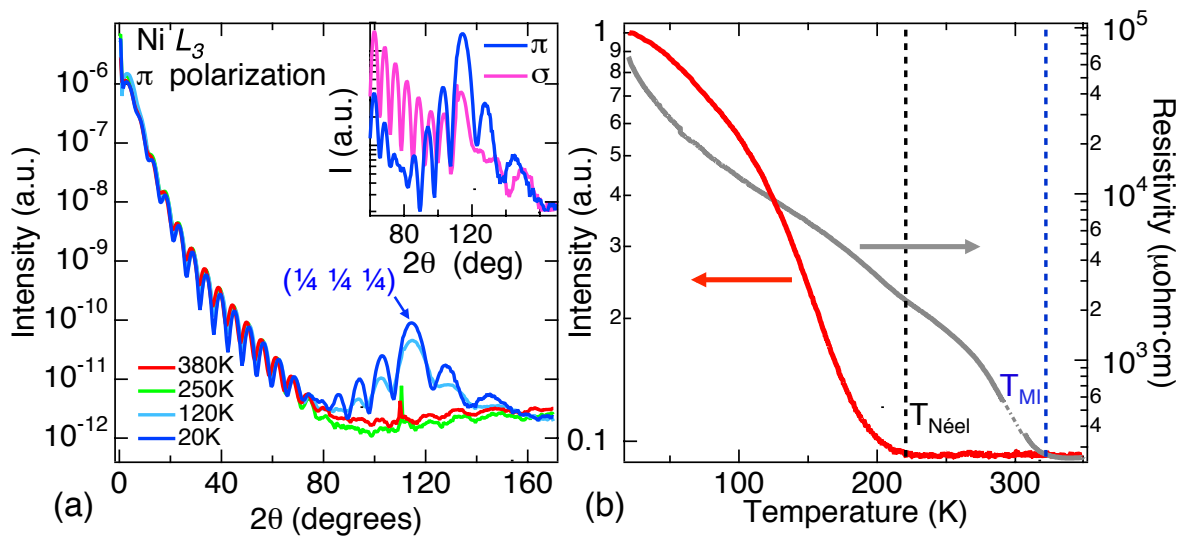

**Figure 3:** Magnetic properties of a 17 nm $(111)_{pc}$ NNO/NGO film probed through resonant soft X-Ray reflectivity. (a) Temperature dependent ω–2θ scans acquired at the Ni $L_3$-edge in π polarized light revealing the appearance of a Bragg reflection with $q_{Bragg} = (¼ ¼ ¼)$ as the temperature is reduced. Inset: polarization dependence of the Bragg peak showing that the superstructure reflection is strongly enhanced for π polarized incoming light. The sharp peak observed at T = 250 K and 380 K is a structural peak from the substrate. (b) Temperature evolution of the peak intensity (red curve, left axis), indicating a Néel transition at $T_{Néel}$ = 230 K. The corresponding resistivity of the film is also displayed (grey curve, right axis).

The magnetic properties of the other $(001)_{pc}$ and $(111)_{pc}$ -oriented NNO films were investigated using similar soft X-ray based techniques, as reported in ref.[23] and ref.[28]. In all cases, we observed a Néel transition at $T_{Néel} = T_{MI}$.

In order to unveil the epitaxial mechanism driving such a remarkable splitting between $T_{MI}$ and $T_{Néel}$, we first note that the impact of strain alone cannot account for the extremely high shift of the MIT. NNO films of comparable thickness subjected to an equivalent or higher level of tensile strain (up to $\varepsilon_{xx}$ = 4.7%) display a 1st order MIT with $T_{MI}$ < 200 K, as confirmed by our transport measurements (Figure 2) and previous studies.[22,24] As can also be seen on Figure 2, the crystallographic orientation on its own does not affect in a corresponding way the (111)$_{pc}$ NNO films deposited on other substrates, such as LAO or LSAT. To try elucidating the observed behavior, we carefully examine both the substrate symmetry and the epitaxial constraints established at the (001)$_{pc}$ and (111)$_{pc}$ substrate/film interfaces, taking into account that the oxygen octahedral connectivity requirements impose additional boundary conditions at the interface. The NGO structure presents an orthorhombic symmetry with a tilt pattern $a^+a^+c^-$ (Glazer notation), with $\Theta$ = 153.2°.[35] In contrast, LSAT is cubic with $\Theta$ = 180°, that corresponds to perfectly aligned $BO_6$ units ($a^0a^0a^0$). Finally, LAO exhibits a rhombohedral distortion and a tilt pattern $a^-a^-a^-$, with $\Theta$ = 174.8°.[36]

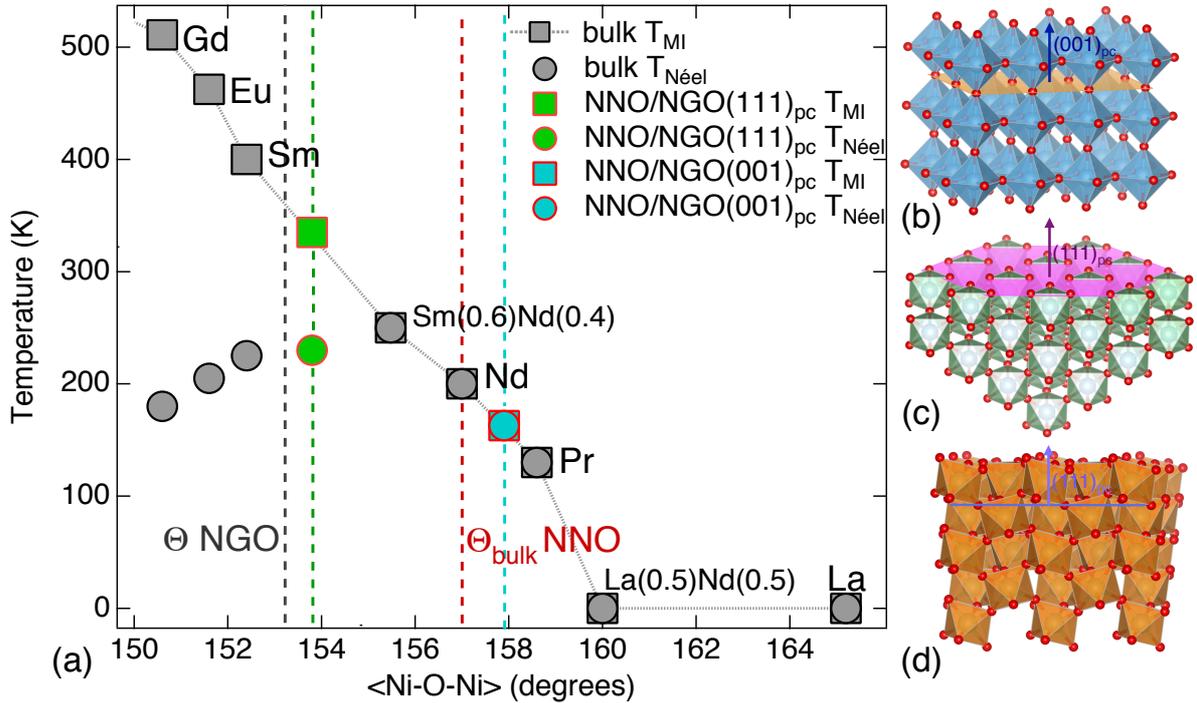

**Figure 4:** Illustration of the epitaxial constraints imposed at the (001)$_{pc}$ and (111)$_{pc}$ film/substrate interfaces and extrapolation of the film Ni-O-Ni bond angle ($\Theta$) from the $R$NiO$_3$ phase diagram.[16,17] (a) The green and blue squares (circles) indicate $T_{MI}$ ($T_{Néel}$) of (111)$_{pc}$ and (001)$_{pc}$ NNO/NGO, respectively, on the phase diagram of bulk $R$NiO$_3$. The green dashed line indicates the extrapolated angle, closer to the tilt of NGO (grey dashed line) than to that of bulk NNO (dashed red line). The blue dashed line points to the angle extrapolated for the case of (001)$_{pc}$ NNO/NGO. (b),(c) Oxygen octahedral network of an ideal cubic perovskite. In (b), along the (001)$_{pc}$ plane, the connectivity between the octahedra is realized through one apical oxygen. In (c), along the (111)$_{pc}$ termination plane (pink), each $BO_6$ unit shares three oxygen sites with the next one. (d) Profile of an orthorhombic perovskite with (111)$_{pc}$ axis oriented out-of-plane. In (b), (c) and (d) the $A$ cation sites are not shown. The bulk data are from ref. [19,37]

The (001)$_{pc}$ substrate/film interface is schematized in Figure 4b. Along the (001)$_{pc}$-direction, each oxygen ochtahedron is connected to the next through a single apical oxygen, forming a rectangular lattice in the plane. In this configuration, the $NiO_6$ units can rather easily adopt the most energetically favorable tilt pattern resulting from the balance between the applied strain and the internal chemical pressure.[38] Our transport properties suggest that the film lattice stabilizes a Ni-O-Ni angle increased with respect to the bulk value ($\Theta$ = 157°), as the MIT of these films is pushed below the bulk $T_{MI}$, independent of the substrate symmetry and of $\varepsilon_{xx}$.[22,24,29]

The scenario dramatically changes when considering the propagation of the substrate rotation patterns along the $(111)_{pc}$ direction (Figure 4c and Figure 4d). In this case, the corner sharing octahedra of the substrate are connected to those of the film through three oxygen atoms, constituting a substantially stronger epitaxial requirement. As a consequence, the $BO_6$ units of the films are forced to rotate and to adopt the tilt angle of the substrate in order to maintain the connectivity across the interface, as can be visualized in Figure 4d. Bearing in mind the $R$-NiO$_3$ phase diagram[16,17], one can observe on Figure 4a that the MIT of $(111)_{pc}$ NNO/NGO at 335K would correspond to $\Theta=153.8°$, an angle which is remarkably close to the tilts characterizing NGO ($\Theta=153.2°$). The observed $T_{Néel}$ is also consistent with that expected for this reduced $\Theta$. The $B$-$O$-$B$ angle inferred from the $T_{MI}$ and $T_{Néel}$ of the film thus lies closer to the $\Theta$ angle of NGO than to the one of bulk NNO. We also note that the impact of the epitaxial constraints at the $(111)_{pc}$ interface is stunningly robust, affecting the whole thickness of a 17 nm thick $(111)_{pc}$ NNO/NGO film.–The behavior of thicker films was also investigated,[28] showing that, as the thickness is increased, lattice relaxation turns on and the films recover their bulk properties.

As an additional test of this picture, we studied the transport properties of NNO films deposited on orthorhombic DyScO$_3$, which provides $\varepsilon_{xx} = 3.9\%$ and $\Theta = 134°$. Again, by choosing an orthorhombic substrate, the $(001)_{pc}$ oriented films display a 1$^{st}$ order MIT below 200 K whereas the $(111)_{pc}$ films reveal an insulating behavior in the whole temperature range investigated (shown in the supplementary information.[28]). It is also interesting to note that the resistivity of $(111)_{pc}$ NNO/NGO in the metallic phase is higher than the resistivity of the other samples, suggesting a smaller *Ni-O* hybridization, as observed in the more distorted compounds.[19,39]

The absence of an analogue orientation driven effect for films deposited on (111) LAO and LSAT substrates is quite surprising, as one may have expected the insulating phase to be completely suppressed, if the *Ni-O-Ni* angle would correspond to the *B-O-B* angle of the substrate. To understand the observed behavior, further structural studies will be necessary in order to provide a detailed description of the mechanisms occurring at the film/substrate interface, taking into account aspects such as lattice mismatch, orientation, symmetry effects, material stiffness, and possible defects.

In summary, we succeeded in extending the phase diagram of NNO films over a temperature range never reported for this compound. Our results unveil the presence of a unique epitaxial effect imposed at the $(111)_{pc}$ interface between the film and the orthorhombically distorted perovskite substrate. In this crystallographic direction, the MIT is found at 335 K, while the transition to the AFM ground state occurs at $T_{Néel}$ = 230 K. These NNO films behave like the $R$NiO$_3$ compounds with $R$-size < Nd. Our work suggests that, by choosing the unconventional crystallographic $(111)_{pc}$ orientation for the growth of NNO perovskite oxide films, it is possible to impose specific tilts of the oxygen octahedra over several atomic layers. These tilts profoundly alter the physical properties of the film, pushing the metal-insulator transition to a temperature that cannot be reached by the application of mere epitaxial strain. Finally, our experimental findings point to a novel remarkable opportunity to tailor perovskite functionalities using heterointerface engineering along the $(111)_{pc}$ orientation.

**Acknowledgments.**

We thank Oleg E. Peil and Antoine Georges for enlightening discussions on the physics of nickelates. We also thank Julien Ruppen, Jérémie Teyssier and Dirk Van der Marel for their stimulating collaboration to the understanding of the material properties. We acknowledge Jennifer Fowlie for proofreading the manuscript and Marco Lopes and Sebastian Muller for their precious technical support. This work was supported by the Swiss National Science Foundation through the National Center of Competence in Research, Materials with Novel Electronic Properties, 'MaNEP' and division II. The research leading to these results has received funding from the European Research Council under the European Union's Seventh Framework Programme (FP7/2007-2013) / ERC Grant Agreement n° 319286 Q-MAC and under grant agreement Nr. 290605 (COFUND: PSI-FELLOW). Part of the research described in this paper was performed at the Canadian Light Source, which is funded by the CFI, NSERC, NRC, CIHR, the Government of Saskatchewan, WD Canada and the University of Saskatchewan.